\documentclass[prd,aps,reprint,floats,nofootinbib,tightenlines,showpacs]{revtex4-1}
\usepackage{epsfig,graphicx,pstricks}
\usepackage{physics}
\usepackage{dsfont}
\usepackage{psfrag}
\usepackage{color}
\usepackage{amsmath}
\usepackage{mathtools}
\usepackage{amsfonts}
\usepackage{amssymb}
\usepackage{textcomp}
\usepackage{revsymb}
\usepackage{multirow}
\usepackage{subfigure}
\usepackage{amsthm}

\usepackage{stackengine}
\stackMath

\begin{document}
	\title {Three-slit interference with a which-path memory ancilla: A bright-dark state formulation}
	\author{Ajay Kumar}
	\author{Guruprasad Kadam }
		\email{guruprasadkadam18@gmail.com}
	\author{Anirban Pathak}
	\email{anirban.pathak@mail.jiit.ac.in}

	\affiliation{Department of Physics and Materials Science and Engineering,\\
		Jaypee Institute of Information Technology,\\
		A-10, Sector-62, Noida, UP-201309.}
	\date{\today}

	\begin{abstract}
	In this paper, we study the effect of a which-path memory ancilla on the three-slit interference pattern within the framework of bright--dark state description [Phys. Rev. Lett. {\bf{134},} 133603 (2025)]. Whereas two slits give one bright mode—a photonic mode that couples to the detector atoms—and one dark mode—a mode that does not couple to the detector atoms—three slits lead to one detector-coupled bright mode and a two-dimensional dark-mode subspace in the three-dimensional path space. We first discuss the classical three-slit interference pattern in terms of probability leakage into the dark subspace. We then discuss the von Neumann entropy ($S_D$) of the dark subspace and the coherence measures: dark-sector coherence ($C_D$) and bright--dark coherence ($C_{BD}$) of the reduced photonic state by tracing out memory, which bring out the internal quantum structure of the two-dimensional dark subspace. We show that, whereas $C_D$  is not a basis independent physical observable by itself, $C_{BD}$ is basis invariant. We establish that for two paths, $C_{BD}$ is fixed entirely by the populations and the single pairwise coherence, while for multi-path $M\geq 3$, it acquires a genuinely multi-path contribution generated by asymmetry among the pairwise which-path overlaps. We finally distinguish the part of the path-coherence loss that is recoverable through measurements of the suitable memory from the irreducible coherence deficit imposed by an uncontrolled environment.
	\end{abstract}
		\maketitle

	\section{Introduction}

Before the advent of quantum theory, two mutually "contradictory" descriptions of the nature of light existed: corpuscular nature chiefly advocated by Newton\cite{Newton1952} and the wave nature attributed to Huygens\cite{Huygens1912}. This dispute over the true nature of light remained unresolved, and  in fact, favoring Newton due to his dominant authority. However, Young's double slit experiment on light diffraction\cite{young1802} conclusively demonstrated the wave nature of light. Subsequent experiment further strengthened this view, and finally received a solid theoretical support when Maxwell unified laws of electricity and magnetic into a set of four beautiful coupled differential equations\cite{Jackson1999}.

Corpuscular theory of light remain dormant until the start of 20th century when Albert Einstein in 1905 explained the photoelectric effect with his light quanta (photon) hypothesis\cite{Einstein1905}, however with considerable skepticism. This was dissolved by Arthur Compton\cite{Compton1923} with his experiments on scattering of X-rays. Quantum theory thus revealed dual description of light;  it sometimes behaves as a wave and other times behave like particle depending on experimental setup. 

Interference is a characteristic wave phenomenon. In the classical description, interference pattern of light is understood as constructive or destructive  interference of classical electromagnetic waves, whereas in quantum theory of Glauber, inspired by the ideas developed by Dirac\cite{Dirac1927a,Dirac1927b},  it is understood as constructive or destructive  interference in terms of superposed transition amplitudes\cite{Glauber1963a,Glauber1963c,Glauber1963b}. In a recent paper Villas-Boas et al.\cite{VillasBoas2025}  formulated quantum description of classical interference pattern in terms of bright and dark states of light which are particular cases of two-mode binomial states. The bright states couple to matter while the dark states do not. This formulation of light-matter interaction in term of bright-dark states (BDS) leads to the result that the destructive interference---points on the screen where the photon-detection probability vanishes--- does not prove that the photons are absent; rather they are in perfect dark states (PDS). Similarly, the constructive interference corresponds to perfectly bright states or maximally superradiant states (MSS). This formulation also predict a new class of states intermediate in nature between PDS and MSS which has no classical counterpart.

The bright–dark framework has found a wide range applications  from constructing a quantum gates\cite{Solak2024}, a novel beam splitter capable of separating a light beam into its two-mode bright and dark components\cite{Solak2025} and explaining continuous-mode diffraction pattern\cite{Cheng2026}, etc. Cheng et al.\cite{Cheng2026} constructed a detector-oriented basis for single slit diffraction, in which a single bright mode is accompanied by an infinite-dimensional dark subspace generated by the continuum of points across the aperture. Few earlier works have developed quantitative duality relations for M-path interferometers, in which the total $l_1$-norm coherence of the reduced path states serve as a generalized visibility and is bounded together with paths distinguishability\cite{Durr2001,Bera2015,Bagan2016,Qureshi2017,Bagan2018,Englert1996}. None of these works, however, addresses the question raised by the intermediate case that we aim to study here: a finite number of discrete paths, $M\geq3$, each carrying its own which-path record. For $M\geq3$ the dark sector at a fixed detector position is a subspace of dimension $M-1$, so it can possess populations, coherence, and entropy of its own. The total $l_1$ coherence used in the duality literature is a single scalar and is blind to this internal structure, while the continuum treatment of Ref.\cite{Cheng2026} does not consider independent which-path markers on countable paths. The internal quantum structure of a finite-dimensional dark sector, and its response to which-path marking and erasure, have not been analyzed in bright-dark formulation to our knowledge.

Three-slit interference has been studied in its own right: Ref.~\cite{Sinha2010} bounded the third-order (Sorkin) interference term\cite{Sorkin1994}, which is identically zero for two paths, and found no deviation from the Born rule.  In Ref.~\cite{Sawant2014} it was theoretically shown using Feynman's path integral approach to quantum mechanics\cite{Feynman1948,FeynmanHibbs1965},  that non-classical looped trajectories can account for the non-zero Sorkin parameter without any modification in standard quantum mechanics. These findings were also experimentally  demonstrated\cite{Rengaraj2018} in the microwave domain.  Ref.~\cite{MaganaLoaiza2016} measured the contribution of   such non-classical looped trajectories  by enhancing the near-fields in the vicinity of the slits through the excitation of surface plasmons. An important point to be noted from all of these works  is that they rely on independent control of the individual slits, which is the level of control required to attach a separate which-path record to each path.  This motivates us to start with the analysis of three-slit  interference pattern within ambit of bright-dark state formulation.   We first construct the collective basis $\{{|B\rangle,|D_1\rangle,|D_2\rangle}\}$, consisting of the detector-coupled bright mode and an orthonormal frame for the two-dimensional dark plane, and show that the classical three-slit intensity pattern, including its principal and secondary maxima, is reproduced by the position-dependent weight of the bright projection, with the complementary probability residing in the dark plane. The distinction between principal maxima, secondary maxima, and minima corresponds to complete, partial, and vanishing alignment, respectively, of the three path phasors with the bright direction.

We then attach a which-path memory ancilla to the paths and study the reduced photonic state. Because the dark sector is two-dimensional, it possesses a nontrivial internal structure whose mixedness can be characterized by the von Neumann entropy $S_D$. A natural further quantity, the $\ell_1$-norm coherence $C_D$ evaluated in the frame $\{|D_1\rangle,|D_2\rangle\}$, turns out not to be basis independent: a unitary rotation within the dark plane leaves the dark subspace itself unchanged but generally alters $C_D$. We therefore introduce the bright--dark coherence $C_{BD}=2|\mathbf{v}|$, where $\mathbf{v}$ is the off-diagonal block of the density matrix connecting the bright direction to the dark plane. This quantity is invariant under unitary rotations of the dark frame because it is anchored to the physically selected bright mode. For two paths, $C_{BD}$ is fixed entirely by the populations and the single pairwise coherence, and therefore carries no information beyond that already contained in the ordinary fringe visibility and phase. For $M\geq3$ paths, it acquires a genuinely multipath character: at the phase-matched detector point and for real overlaps, it vanishes when all pairwise which-path overlaps are equal, while asymmetric pairwise overlaps can generate a nonzero $C_{BD}$. Thus, in the multipath case, $C_{BD}$ provides a basis-independent probe of asymmetry among the which-path records that is not resolved by a single visibility measure.

Finally, we ask what kinds of dark-sector mixedness can arise and how they can be distinguished. We consider a photon coupled both to a controlled memory, which can be measured, and to an uncontrolled environment, which cannot. The reduced photonic state depends on the two sets of overlaps, $\mu_{ij}$ and $\eta_{ij}$ between memory states $\{\ket{m_i}\}$ and  environment states $\{\ket{e_i}\}$ respectively, only through their products, so no functional of the reduced state---including $S_D$, $C_{BD}$, and the dark-sector purity---can determine how the mixedness is divided between the memory and the environment. A projective measurement of the memory followed by postselection can resolve this ambiguity. To quantify the recoverability of photonic coherence through such a memory measurement, we define the probability-weighted average conditional path coherence $\overline{\mathcal{C}}{\mathrm{path}}$. It is bounded from below by the unconditional path coherence $\mathcal{C}{\mathrm{path}}(\rho_p)$ of the reduced photonic state $\rho_p$ and from above by the optimized conditional path coherence $\mathcal{C}_{\mathrm{path}}^{\mathrm{opt}}$, obtained by optimizing over the choice of memory basis. In particular, a suitably chosen memory basis can make the measurement outcomes carry no which-path information while transferring the path dependence into conditional phase relations

We organize the paper as follows: We discuss bright-dark state formulation of two-slits in Sec.\ref{sec:BDS-2slit} and then discuss which-path marker effect on the bright-dark states in Sec.\ref{sec:whichpath-2slit}. We then discuss three-slit extenstion and its application to a single photon interference pattern in Sec.\ref{sec:BDS_3-slit}. In Sec.\ref{sec:entr-coherence},  we discuss dark-subspace entropy and coherence measures. Sec.\ref{sec:eraser} discuses the quantum erasure within ambit of bright-dark state formulation. Finally, we conclude in Sec.\ref{sec:conclusion}.

	\section{Bright and dark states: two slits}
	\label{sec:BDS-2slit}
	In this section we recapitulate the bright-dark state formulation of Villas-Boas, et al\cite{VillasBoas2025}. According to quantum theory\cite{Glauber1963a,Glauber1963b,Glauber1963c,Sudarshan1963}, the statistical properties of a given field can be derived from its electric field operator
	
	\begin{equation}
		\mathbf{E}(\mathbf{r},t)=\mathbf{E}^{(+)}(\mathbf{r},t)+\mathbf{E}^{(-)}(\mathbf{r},t),
		\label{efield_op}
	\end{equation}
	where $\mathbf{E}^{(+)}\propto a$ and $\mathbf{E}^{(-)}\propto a^\dagger$, where $a, a^\dagger $ are the annihilation and creation operator respectively. The probability that a photon in a single-mode state $\ket{\Psi}$ is absorbed by a sensor atom is proportional to
	\begin{equation}
		\bra{\Psi}\mathbf{E}^{(-)}\mathbf{E}^{(+)}\ket{\Psi}
		\propto
		\bra{\Psi}a^{\dagger}a\ket{\Psi}.
	\end{equation}
	This expression follows from the energy-exchange interaction between the field and the sensor, described in the rotating-wave approximation by
	\begin{equation}
		H=\mathbf{E}^{(+)}(\mathbf{r},t)\,\sigma_{+}+\mathbf{E}^{(-)}(\mathbf{r},t)\,\sigma_{-},
	\end{equation}
	where $\sigma_{+}$ and $\sigma_{-}$ are the raising and lowering operators of the sensor atom. For a single mode, Eq.~(3) shows that the excitation probability is proportional to $\langle a^{\dagger}a\rangle$, namely the mean photon number. Therefore only zero-intensity fields, for which $\langle a^{\dagger}a\rangle=0$, are unable to excite the sensor atom; in the single-mode case this is just the vacuum state.
	
	For the two-mode case, let $a$ and $b$ denote the annihilation operators  and let $\theta$ be their relative phase. Then the positive-frequency field operator takes the form
	\begin{equation}
		\mathbf{E}^{(+)}(\mathbf{r},t)\propto a+b\,e^{i\theta}.
	\end{equation}
	The corresponding sensor-excitation probability is proportional to
	\begin{equation}
		\bra{\Psi}\mathbf{E}^{(-)}\mathbf{E}^{(+)}\ket{\Psi}
		\propto
		\bra{\Psi}\left(a^{\dagger}+b^{\dagger}e^{-i\theta}\right)\left(a+b\,e^{i\theta}\right)\ket{\Psi}.
	\end{equation}
	Accordingly, the interaction Hamiltonian becomes
	\begin{equation}
		H=g\left(a+b\,e^{i\theta}\right)\sigma_{+}+g\left(a^{\dagger}+b^{\dagger}e^{-i\theta}\right)\sigma_{-},
	\end{equation}
	where $g=g(\mathbf{r})$ is the atom--field coupling strength which we assume to be equal following \cite{VillasBoas2025}. It is then natural to introduce the collective operators\cite{VillasBoas2025}
	\begin{equation}
		c=\frac{a+b\,e^{i\theta}}{\sqrt{2}},
		\qquad
		d=\frac{-a\,e^{-i\theta}+b}{\sqrt{2}}.
	\end{equation}
	The state with N excitation can be written in terms of collective basis ($\{c,d\}$):
\begin{equation}
	\ket{\psi_n^N(\theta)} \equiv \ket{n, N-n}_{c,d}
	= \frac{(c^\dagger)^n (d^\dagger)^{N-n}}{\sqrt{\, n!\,(N-n)!}} \ket{0,0}_{c,d},
\end{equation}
	or the bare basis  ($\{a,b\}$):
	\begin{equation}
		\ket{\psi_n^N(\theta)} = \frac{(a^\dagger + b^\dagger e^{-i\theta})^n (-a^\dagger e^{i\theta} + b^\dagger)^{N-n}}{\sqrt{2^N\, n!\,(N-n)!}} \ket{0,0}_{a,b}.
	\end{equation}
	In the collective Fock basis the state with $n=0$ is annihilated by  $\mathbf{E}^{(+)}(\propto c)$.  These are the states with no bright-mode quanta and they do not excite the sensor atom. They are termed perfectly dark states (PDS)\cite{VillasBoas2025}. Written explicitly in the bare two-mode basis,
	\begin{equation}
		\ket{\psi^{N}_{0}(\theta)}=
		\sqrt{\frac{N!}{2^N}}
		\sum_{m=0}^{N}
		\frac{(-1)^m e^{i m \theta}}{\sqrt{m!(N-m)!}}\,
		\ket{m, N-m}_{a,b}.
	\end{equation}
	Conversely, the states with all excitations in the bright mode, $\ket{N,0}_{c,d}$, form the perfectly bright states/maximally superadiant states (MSS)\cite{Dicke1954}:
	\begin{equation}
		\ket{\psi^{N}_{N}(\theta)}=e^{-iN\theta}
		\sqrt{\frac{N!}{2^N}}
		\sum_{m=0}^{N}
		\frac{e^{i m \theta}}{\sqrt{m!(N-m)!}}\,
		\ket{m, N-m}_{a,b}.
	\end{equation}
For a single photon incident on a double-slit, one may write
	\begin{equation}
		\ket{S}=\frac{1}{\sqrt{2}}\left(e^{-i\mathbf{k}_1\cdot \mathbf{d}_1}\ket{1,0}_{a,b}+e^{-i\mathbf{k}_2\cdot \mathbf{d}_2}\ket{0,1}_{a,b}\right).
	\end{equation}
	At a detector position characterized by the relative phase $\theta=(\mathbf{k}_2-\mathbf{k}_1)\cdot \mathbf{r}$, the one-photon bright and dark states are
	\begin{equation}
		\ket{\psi^{1}_{1}(\theta)}=\frac{1}{\sqrt{2}}\left(\ket{1,0}_{a,b}+e^{-i\theta}\ket{0,1}_{a,b}\right),
\end{equation}
\begin{equation}
		\ket{\psi^{1}_{0}(\theta)}=\frac{1}{\sqrt{2}}\left(\ket{1,0}_{a,b}-e^{-i\theta}\ket{0,1}_{a,b}\right).
	\end{equation}
	Using these states, the single-photon field can be re-written as
	\begin{equation}
		\ket{S'}=\cos\!\left(\frac{\delta\phi}{2}\right)\ket{\psi^{1}_{1}(\theta)}-i\sin\!\left(\frac{\delta\phi}{2}\right)\ket{\psi^{1}_{0}(\theta)},
	\end{equation}
	where $\delta\phi=-(\mathbf{k}_2\cdot \mathbf{d}_2-\mathbf{k}_1\cdot \mathbf{d}_1)+\theta=\mathbf{k}_2\cdot \mathbf{r}_2-\mathbf{k}_1\cdot \mathbf{r}_1$ is the path-phase difference between the two slits and the detector. Thus, according to bright-dark states (BDS) formulation, the photon can be present at any detector position, but the sensor responds only to the bright component, while the dark component remains undetected.
	
	\section{Which-path  memory ancilla and the purity of bright/dark states}
	\label{sec:whichpath-2slit}
	We now implements the standard which-path marking mechanism, in which distinguishable auxiliary records suppress path interference through entanglement with the interfering system\cite{ScullyDruhl1982, Scully1991}. We  attach a controlled memory state $\ket{m_i}$\cite{Walborn2002} and an environmental state $\ket{e_i}$ to the $i^{\text{th}}$ path. The marked single-photon state is then
	\begin{equation}
		\ket{\Psi_{ME}}=\frac{1}{\sqrt{2}}\left(\ket{1,0}_{a,b}\ket{m_1}\ket{e_1}+e^{-i\delta\phi}\ket{0,1}_{a,b}\ket{m_2}\ket{e_2}\right),
		\label{2-slit-mem-envi-state}
	\end{equation}
	with
	\begin{equation}
		\mu\equiv\braket{m_1}{m_2},
		\qquad
		\eta\equiv\braket{e_1}{e_2}.
		\label{mu-eta}
	\end{equation}
	Here $\mu$ quantifies the coherence left after the controlled which-path memory is ignored, while $\eta$ quantifies the coherence left after ignoring the uncontrolled environment. Tracing out both the memory and the environment gives
\begin{align}
	\rho_p
	&=\frac{1}{2}\Big(
	\ket{1,0}\bra{1,0}
	+\ket{0,1}\bra{0,1} \nonumber\\
	&\qquad
	+\chi^* e^{i\delta\phi}\ket{1,0}\bra{0,1}
	+\chi e^{-i\delta\phi}\ket{0,1}\bra{1,0}
	\Big),
	\label{rho-two-reduced}
\end{align}

where $\chi\equiv\mu\eta$.
	
	Consider a dark fringe of the unmarked experiment, $\delta\phi=(2n+1)\pi$. In the detector bright--dark basis, Eq.(\ref{rho-two-reduced})  becomes
	
	\begin{equation}
		\begin{aligned}
			\rho_p^{\rm dark}={}&\frac{1-\Re\chi}{2}\ket{\psi_1^1(\theta)}\bra{\psi_1^1(\theta)}+\frac{1+\Re\chi}{2}\ket{\psi_0^1(\theta)}\bra{\psi_0^1(\theta)}\\
			&+\frac{i\Im\chi}{2}\Big(\ket{\psi_1^1(\theta)}\bra{\psi_0^1(\theta)}-\ket{\psi_0^1(\theta)}\bra{\psi_1^1(\theta)}\Big).
		\end{aligned}
		\label{rho-dark-basis}
	\end{equation}
	If neither the memory nor the environment stores which-path information, identical records give $\mu=\eta=1$, hence $\chi=1$, and one has
	\begin{equation}
		\rho_p^{\rm dark}=\ket{\psi_0^1(\theta)}\bra{\psi_0^1(\theta)},
	\end{equation}
	which is the pure dark state. By contrast, once either the memory or the environment carries partial path information, $|\chi|=|\mu\eta|<1$ and the photonic state is mixed, with purity
	\begin{equation}
		\mathrm{Tr}\!\left[(\rho_p^{\rm dark})^2\right]=\frac{1+|\chi|^2}{2}=\frac{1+|\mu\eta|^2}{2}<1.
		\label{purity-dark}
	\end{equation}
	
The amount of bright-state contamination inside a dark-fringe (i.e, state at a location that would have otherwise been dark in the absence of marker now contains some detector-coupled bright component) is given by the bright weight
\begin{equation}
 W_B^{\text{dark}}=\text{tr}(\rho_B \rho_p^{\text{dark}}),
 \end{equation}
 where $\rho_B=\ket{\psi^1_1(\theta)}\bra{\psi_1^1(\theta)}$. With $\rho_p^{\text{dark}}$ given by (\ref{rho-dark-basis}) we get 
	
	\begin{equation}
		W_B^{\rm dark}=\bra{\psi_1^1(\theta)}\rho_p^{\rm dark}\ket{\psi_1^1(\theta)}=\frac{1-\Re\chi}{2}=\frac{1-\Re(\mu\eta)}{2},
		\label{bright-weight-dark}
	\end{equation}
	which is zero only for the pure dark state ($\chi=1$). Thus the dark fringe is washed out leading to reduced interference visibility precisely because the path information is stored  either in the controlled memory ancilla $\ket{m_i}$ or the uncontrolled environment $\ket{e_i}$ which destroys the purity of the dark state (Eq.(\ref{purity-dark})) and populate a nonzero bright component. References \cite{Herzog1995, Kwiat1992} gives  experimental demonstration of the loss and revival of the interference visibility with entangled photons.

	\section{Dark-bright state formulation for the three slits}
\label{sec:BDS_3-slit}

\subsection{Collective operators and states}
	In case of three slits we have three modes: $a_i, i=1,2,3$.	The detector couples to the symmetric combination of all three modes. With relative phases $\theta_2$ (mode $a_2$) and $\theta_3$ (mode $a_3$) measured against mode $a_1$, the positive-frequency operator is

	\begin{equation}
		E^{(+)} \propto a_1 + a_2\,e^{i\theta_2} + a_3\,e^{i\theta_3}.
	\end{equation}

	The (normalized) collective operators are: 
	\begin{equation}
		B = \frac{1}{\sqrt{3}}\left(a _1+ a_2\,e^{i\theta_2} + a_3\,e^{i\theta_3}\right),
	\end{equation}
	while the dark sector is spanned by two operators $D_1$ and $D_2$
	orthogonal to $B$ given by
	\begin{equation}
		D_1=\frac{1}{\sqrt{2}}\left(a_1-e^{i\theta_2}a_2\right),
	\end{equation}
	\begin{equation}
		D_2=\frac{1}{\sqrt{6}}\left(a_1+e^{i\theta_2}a_2-2e^{i\theta_3}a_3\right).
	\end{equation}
	
	The general three-slit Fock sate with N-excitations in the collective basis ($\{B,D_1,D_2\}$) is
	
\begin{align}
	\ket{\psi^N_{n_B, n_{D_1},n_{D_2}}}
	&\equiv \ket{n_B,n_{D_1},n_{D_2}}_{B,D_1,D_2}\nonumber\\
	&=\frac{(B^\dagger)^{n_B}(D_1^\dagger)^{n_{D_1}}(D_2^\dagger)^{n_{D_2}}}{\sqrt{n_B!\,n_{D_1}!\,n_{D_2}!}}\ket{0,0,0}_{B,D_1,D_2},
\end{align}
with $N=n_B+n_{D_1}+n_{D_2}$.
	
Maximally superradiant states (MSS) corresponds to $n_{D_1}=0, n_{D_2}=0$:

	\begin{equation}
		\ket{\psi^{N}_{0,0}} = \frac{(B^{\dagger})^{N}}{\sqrt{N!}}\,\ket{0,0,0},
	\end{equation}
	which, expanded in the bare basis, reads
	\begin{align}
		\ket{\psi^{N}_{N, 0, 0}}
		&= \sqrt{\frac{N!}{3^{N}}}
		\sum_{m_1,m_2,m_3}
		\frac{e^{-i(m_2\theta_2 + m_3\theta_3)}}
		{\sqrt{m_1!\,m_2!\,m_3!}}\nonumber\\
			&\hspace{4.0em}
		\times\ket{m_1,m_2,m_3}_{a_1,a_2,a_3},
	\end{align}
		with $m_1+m_2+m_3=N$.
		
	For three slits the perfectly dark state (PDS) is no longer unique: it becomes
	an $(N+1)$-dimensional subspace, spanned by distributing the $N$ photons
	between the two dark operators,
	\begin{equation}
		\ket{\psi^{N}_{0, k}}
		= \frac{(D_1^{\dagger})^{k}\,(D_2^{\dagger})^{N-k}}
		{\sqrt{k!\,(N-k)!}}\,\ket{0,0,0}_{B,D_1,D_2},
	\end{equation}
	where $ k = 0,1,\dots,N$, and each of which satisfies
	\begin{equation}
		E^{(+)}\,\ket{\psi^{N}_{k}} = 0.
	\end{equation}
In bare basis, a general PDS can be written as
	\begin{align}
		\ket{\mathrm{PDS}^{N}}
		&= \sum_{m_1,m_2,m_3}
		\Lambda_{m_1 m_2 m_3}(\theta_2,\theta_3)\nonumber\\
	 &\hspace{5.0em}
		\times\ket{m_1,m_2,m_3}_{a_1,a_2,a_3},
	\end{align}
	with $m_1+m_2+m_3=N$. The coefficients $\Lambda_{m_1 m_2 m_3}$ are fixed by orthogonality
	to $B$ but depend on the choice of basis within the dark subspace.
	
	\subsection{Single Photon Three-slit Interference}
	 For a single photon impinging on three-slit it is straightforward to obtain analog of the two-slit decomposition $\ket{S'}$: 
		\begin{equation}
		\ket{S_{3-\text{slit}}'}=c_B
		\ket{B}
		+c_{D_1}\ket{D_1}
		+c_{D_2}\ket{D_2}.
	\end{equation}

	where,
	
\begin{align}
	c_B
	&=\frac{1+e^{-i\alpha_2}+e^{-i\alpha_3}}{3},\qquad
	&c_{D_1}=\frac{1-e^{-i\alpha_2}}{\sqrt6},\nonumber\\
	&c_{D_2}=\frac{1+e^{-i\alpha_2}-2e^{-i\alpha_3}}{\sqrt{18}}.
\end{align}
where $\alpha_i=\delta_{i}-\theta_i$ and $\delta_{i}$ is the path phase  of the incident photon.
	Since the sensor couples only to the bright mode, the entire three-slit
	pattern is the position dependence of the bright projection
	$P_{\mathrm{bright}}=|c_B|^{2}$, with the complementary weight
	$P_{\mathrm{dark}}=|c_{D_1}|^{2}+|c_{D_2}|^{2}=1-P_{\mathrm{bright}}$ leaking
	into the two-dimensional dark plane. The bright weight
\begin{equation}
	P_{\rm bright}
	=|c_B|^2
	=\frac{1}{9}
	\left|1+e^{i\alpha_2}+e^{i\alpha_3}\right|^2.
	\label{Pdet-phasor}
\end{equation}
	makes the structure of the interference pattern transparent: each maximum is a position
	where the three phasors reinforce, and its height is set by how completely
	they align with the bright direction $\ket{B}$. For equally spaced slits
	($\alpha_2=\phi$, $\alpha_3=2\phi$) Eq.~\eqref{Pdet-phasor} reduces to the well
	known classical three-slit interference form
	\begin{equation}
		P_{\mathrm{3-slit}}(\phi)=\frac{1}{9}
		\left[\frac{\sin(3\phi/2)}{\sin(\phi/2)}\right]^{2}.
		\label{grating}
	\end{equation}
	
\begin{figure}[h]
		\vspace{-0.4cm}
		\begin{center}
			\includegraphics[width=9cm,height=5cm]{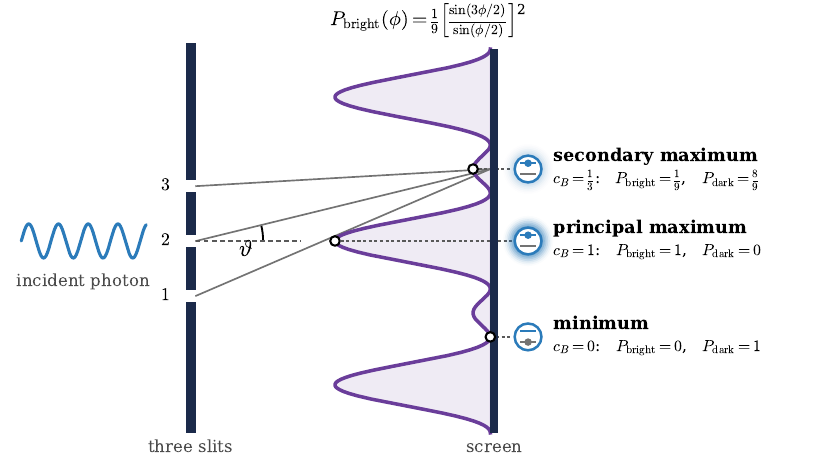}
			\caption{Figure illustrating single photon three-slit interference pattern \eqref{grating}. Note that the angular position $\vartheta$ is related to $\phi$: $\phi=\frac{2\pi d \text{sin}\vartheta}{\lambda}$, where $d$ is the distance between successive slits and $\lambda$ is the wavelength of light.  } 
			\label{3-slit}
		\end{center}
	\end{figure}

	Fig.\ref{3-slit} shows single photon three-slit interference pattern described by \eqref{grating}. The interference pattern is characterized by three distinct locations: principle maxima, minima and secondary maxima.
	
	\paragraph{Principal maxima.}
	When all three phasors are aligned, $\alpha_2=\alpha_3=0\ (\mathrm{mod}\,2\pi)$, the slit amplitudes add up to exactly the symmetric combination the detector couples to,

	\begin{equation}
		c_B=\frac{1+1+1}{3}=1,
		\qquad P_{\mathrm{bright}}=1,
		\qquad P_{\mathrm{dark}}=0 .
	\end{equation}
	
	The photon is then entirely in the bright state $\ket{B}$, with no orthogonal component. Principal maxima reach the full height because the bright direction is the in-phase symmetric state, so perfect in-phase addition is
	identical to complete projection onto $\ket{B}$ with zero leakage into dark subspace.
	
	\paragraph{Secondary maximum.}
	The single secondary maximum per period occurs at $\alpha_2=\pi$,
	$\alpha_3=2\pi$ (i.e.\ $\phi=\pi$), where the phasors are $1,\,-1,\,+1$ and
	sum to $1$,

	\begin{equation}
		c_B=\frac{1}{3},
		\qquad P_{\mathrm{bright}}=\frac{1}{9},
		\qquad P_{\mathrm{dark}}=\frac{8}{9}.
	\end{equation}

	This is a local maximum of the bright projection, but a weak one: two phasors
	reinforce while the third opposes, leaving a residual bright amplitude of
	magnitude $1/3$, so that the remaining $8/9$ of the photon has leaked into the
	dark plane. The height ratio of the secondary to the principal maximum,
	$(1/3)^{2}=1/9$, is precisely the squared ratio of residual to full bright
	amplitude. We stress that the secondary maximum is a feature of the bright
	\emph{weight}, not of any change in the bright state itself: the bright
	projection always points along the pure direction $\ket{B}$, and what varies
	across the screen is how much amplitude reaches it.
	
	Interestingly, the secondary maxima provides a natural 3-path realization of the intermediate bright-dark states identified in \cite{VillasBoas2025}. Although it is local intensity maximum, the corresponding state is predominantly dark.
	
	\paragraph{Minima.}
	Between the maxima the pattern vanishes at $\alpha=2\pi/3$ and $4\pi/3$,
	where the three phasors sum to zero,

	\begin{equation}
		c_B=0,\qquad P_{\mathrm{bright}}=0,\qquad P_{\mathrm{dark}}=1 .
	\end{equation}

	The photon is then entirely in the dark plane; it is fully present but unable to excite the
	the sensor. Thus vanishing detection probability does not imply the absence of photons.
	
Thus, in the bright-dark-states formulation, the interference pattern can be explained in terms of  a continuous exchange of
probability between a single bright direction and the dark hyperplane: principal maxima are points of complete bright projection (perfect phasor alignment, zero dark-plane leakage), the secondary maximum is a point of partial alignment where a weak bright point suffers a substantial leakage into dark-plane, and minima are points of complete dark leakage. The appearance of two distinct types of maxima: principle and secondary, reflects the fact that three phasors admit partial-alignment configurations.

	\section{Dark-subspace entropy and coherence}
	\label{sec:entr-coherence}
	We have seen that the  central new feature of the three-slit bright--dark formulation is that the dark states form a two-dimensional subspace. Consequently, the dark sector itself can possess coherence, entropy, and internal quantum structure independently of the detector-coupled bright component. This allows one to distinguish between different kinds of darkness that are indistinguishable at the level of ordinary intensity measurements.
	
	\subsection{Von Neumann entropy}
	
For a single photon impinging on three-slits the orthonormal basis is
	\begin{equation}
		\{|B\rangle, |D_1\rangle, |D_2\rangle\},
	\end{equation}
	with
	\begin{equation}
		\langle B|D_i\rangle = 0,
		\qquad
		\langle D_i|D_j\rangle=\delta_{ij}.
	\end{equation}
	
	The projector onto the dark subspace is
	\begin{equation}
		P_D	= |D_1\rangle\langle D_1|	+|D_2\rangle\langle D_2|.
		\label{PD}
	\end{equation}
	
	For a general reduced photonic density matrix $\rho_p$, the dark-sector density matrix is
	\begin{equation}
		\rho_p = \frac{P_D \rho_p P_D}
		{\mathrm{Tr}(P_D\rho_p)}.
		\label{rhoD}
	\end{equation}
	
The dark-sector weight is
	\begin{equation}
		W_D	= \mathrm{Tr}(P_D\rho_p)
		\label{WD}
	\end{equation}
	Unlike the detector-coupled bright weight ($W_B$), which determines the observed screen intensity, $\rho_D$ characterizes the internal structure of the  component of the field which is blind to the detector.

	To quantify the internal structure of the dark sector, we define the dark-subspace von Neumann entropy
	\begin{equation}
		S_D=-\mathrm{Tr}(\rho_D \ln \rho_D).
		\label{SD}
	\end{equation}
	
A pure dark state has the form
	\begin{equation}
		|\psi_D\rangle
		=
		\alpha |D_1\rangle + \beta |D_2\rangle,
		\qquad
		|\alpha|^2+|\beta|^2=1,
		\label{pure_dark}
	\end{equation}
	with density matrix
	\begin{equation}
		\rho_D^{(\mathrm{pure})}
		=
		|\psi_D\rangle\langle\psi_D|.
		\label{rho_pure}
	\end{equation}
	
	The von Neumann entropy of the pure state is
	\begin{equation}
		S_D = 0,
		\label{SD_pure}
	\end{equation}
	
		By contrast, a mixed dark state may be written as
	\begin{equation}
		\rho_D^{(\mathrm{mix})}
		=
		p_1 |D_1\rangle\langle D_1|
		+
		p_2 |D_2\rangle\langle D_2|,
		\label{rho_mix}
	\end{equation}
	where $p_1+p_2=1$.    For a maximally mixed dark state ($p_1=p_2=\frac{1}{2}$),
	\begin{equation}
		\rho_D
		=
		\frac12
		\left(
		|D_1\rangle\langle D_1|
		+
		|D_2\rangle\langle D_2|
		\right),
		\label{maxmix}
	\end{equation}
	one obtains
	\begin{equation}
		S_D = \ln 2.
		\label{SD_max}
	\end{equation}
	
	Thus the dark-subspace entropy measures the degree of mixedness within the detector-invisible sector. Physically, $S_D$ quantifies how organized the dark sector is. A pure dark state corresponds to a coherent cancellation against the detector bright mode, whereas a mixed dark state corresponds to a statistical distribution over different dark directions.
	
	\subsection{Dark-sector coherence}
	
	In addition to entropy, one may define a dark-sector coherence measure\cite{Baumgratz2014,Streltsov2017}
	\begin{equation}
		\mathcal C_D
		=
		\sum_{i\neq j}
		|\langle D_i|\rho_D|D_j\rangle|.
		\label{CD}
	\end{equation}
	
	For the pure state of Eq.~(\ref{pure_dark}),
	\begin{equation}
		\mathcal C_D
		=
		2|\alpha\beta|,
		\label{CD_pure}
	\end{equation}
	which is maximal for equal superpositions and vanishes when the state occupies a single dark basis state.
	
	For the mixed state of Eq.~(\ref{rho_mix}),
	\begin{equation}
		\mathcal C_D = 0.
		\label{CD_mix}
	\end{equation}
	
	This quantity therefore measures coherence internal to the dark manifold rather than ordinary interference visibility.	Note that the darkness at a detector point and coherence are distinct physical properties; a state may remain completely dark while still possessing substantial hidden coherence inside the dark subspace.
	
	\subsection{Which-path ancilla and dark-sector mixing}
	 The marked single-photon state is written as
	 
	\begin{equation}
		\lvert\Psi_{M}\rangle=\tfrac{1}{\sqrt3}\sum_i e^{-i\delta_i}
	\lvert i\rangle\lvert m_i\rangle
	\label{marked-state}
	\end{equation}
	(with $\delta_1=0$). Tracing out memory, we get reduced photon state

	\begin{equation}
		\rho_p
		=
		\frac13
		\sum_i |i\rangle\langle i|
		+
		\frac13
		\sum_{i\neq j}
		\Gamma_{ij}|i\rangle\langle j|,
		\label{rhop_general}
	\end{equation}
	where,
	
	\begin{equation}
		\Gamma_{ij}=\mu_{ji}e^{-i(\delta_i-\delta_j)},\quad
\mu_{ji}=\langle m_j\lvert m_i\rangle
\end{equation}

	The overlaps $\mu_{ij}$ therefore determine not only the bright-state visibility but also the internal structure of the dark sector after projection with $P_D$.
	
	For the three-slit case, the dark-sector density matrix is the normalized two-dimensional block of $\rho_p$ in the detector-adapted basis $\{\ket{D_1},\ket{D_2}\}$:
	\begin{equation}
		\rho_p^{\rm dark}
		=
		\frac{P_D\rho_p P_D}{\Tr(P_D\rho_p)}
		=
		\frac{1}{W_D}
		\begin{pmatrix}
			A_{11} & A_{12}\\[4pt]
			A_{12}^* & A_{22}
		\end{pmatrix}_{D_1,D_2},
		\label{rho-three-dark}
	\end{equation}
	where
	\begin{equation}
		W_D=A_{11}+A_{22}=1-W_B
		\label{WD-three-explicit}
	\end{equation}
	is the total dark-sector weight. Using the dark states defined above, the matrix elements are
	\begin{equation}
		A_{11}=	\frac13	-
		\frac13\Re\!\left(\Gamma_{12}e^{-i\theta_2}\right),
		\label{A11-three-dark}
	\end{equation}
	\begin{equation}
		A_{22}=\frac13	+	\frac19\Re\!\left(
		\Gamma_{12}e^{-i\theta_2}
		-2\Gamma_{13}e^{-i\theta_3}
		-2\Gamma_{23}e^{-i(\theta_3-\theta_2)}
		\right),
		\label{A22-three-dark}
	\end{equation}
	and
	\begin{align}
		A_{12}
		&=
		\frac{1}{3\sqrt{12}}
		\Bigl(
		\Gamma_{12}e^{-i\theta_2}
		-\Gamma_{12}^{*}e^{i\theta_2}
		-2\Gamma_{13}e^{-i\theta_3}
		\nonumber\\
		&\qquad\qquad
		+2\Gamma_{23}e^{-i(\theta_3-\theta_2)}
		\Bigr).
		\label{A12-three-dark}
	\end{align}
	Thus, unlike the two-slit case, $\rho_p^{\rm dark}$ is generally a $2\times2$ density matrix rather than a single number multiplying one dark state. Its diagonal entries give the populations of the two dark directions, while $A_{12}$ gives the coherence hidden inside the detector-invisible dark manifold.
	
	As the ancilla states become more distinguishable, the off-diagonal coherence in $\rho_D$ are suppressed, causing
	\begin{equation}
		\mathcal C_D \rightarrow 0,
		\qquad
		S_D \rightarrow \ln 2.
		\label{limits}
	\end{equation}
	
	Thus controlled which-path memory transforms a coherent dark superposition into a mixed dark manifold.
	
	\subsection{Bright-dark sector coherence}
	\label{sec:BD_coherence}
However, the dark-sector coherence $C_D$ of Eq.~(53) is not a basis independent physical observable.
	It is an $\ell_1$-norm coherence evaluated in the frame $\{\lvert D_1\rangle,
	\lvert D_2\rangle\}$, but within the two-dimensional dark plane there is no
	preferred frame: any unitary $U\in U(2)$ acting inside the plane,
	$\lvert D_a\rangle \to \sum_b U_{ab}\lvert D_b\rangle$, leaves the detector
	response, the dark-sector weight $W_D$, and the dark-subspace entropy $S_D$
	unchanged, yet alters $C_D$ continuously. We therefore replace $C_D$ by a coherence anchored to the detector-coupled bright mode
	$\lvert B\rangle$.  
	
	The reduced single-photon state in the detector-adapted basis
	$\{\lvert B\rangle,\lvert D_1\rangle,\lvert D_2\rangle\}$ in block form,
	\begin{equation}
		\rho_p=
		\begin{pmatrix}
			W_B & \mathbf{v}^{\dagger}\\[2pt]
			\mathbf{v} & \rho_{DD}
		\end{pmatrix},
		\qquad
		\mathbf{v}=
		\begin{pmatrix}
			\langle D_1\lvert\rho_p\rvert B\rangle\\[2pt]
			\langle D_2\lvert\rho_p\rvert B\rangle
		\end{pmatrix},
		\label{rhoblock}
	\end{equation}
	where $W_B=\langle B\lvert\rho_p\rvert B\rangle$ is the bright weight,
	$\rho_{DD}$ is the (unnormalized) $2\times2$ dark block with
	$\mathrm{Tr}\,\rho_{DD}=W_D=1-W_B$, and $\mathbf{v}$ is the bright--dark
	off-diagonal block. A frame rotation inside the dark plane acts as
	$\mathbf{v}\to U\mathbf{v}$ and $\rho_{DD}\to U\rho_{DD}U^{\dagger}$, so while
	the off-diagonal element of $\rho_{DD}$ (the object behind $C_D$) is frame
	dependent, the norm $\lvert\mathbf{v}\rvert^2=\mathbf{v}^{\dagger}\mathbf{v}$ is
	invariant. 
	
	We define the bright--dark coherence
	\begin{equation}
			C_{BD}^{\text{3-slit}}=2\lvert\mathbf{v}\rvert
			=2\sqrt{\lvert\langle B\lvert\rho_p\rvert D_1\rangle\rvert^{2}
				+\lvert\langle B\lvert\rho_p\rvert D_2\rangle\rvert^{2}}
		\label{CBDdef}
	\end{equation}
	$C_{BD}$ is invariant under every $U(2)$ rotation of the dark frame and is fixed
	only by the detector-selected direction $\lvert B\rangle$ and the state.
	
 Because
	$\langle D_a\lvert\openone\rvert B\rangle=\langle D_a\lvert B\rangle=0$, the
	identity part of $\rho_p$ drops out of $\mathbf{v}$ entirely: $C_{BD}$ is built
	solely from the memory overlaps. With the phase mismatch $\alpha_i\equiv\delta_i-\theta_i$ between source and detector phases, a straightforward calculation gives
	\begin{align}
		\langle D_1\lvert\rho_p\rvert B\rangle
		&=\frac{\sqrt6}{18}\,e^{-i\alpha_2}\!
		\Big(\mu_{12}^{*}e^{2i\alpha_2}+\mu_{13}^{*}e^{i(\alpha_2+\alpha_3)}
		\notag\\[-2pt]
		&\hspace{4.2em}-\mu_{23}^{*}e^{i\alpha_3}-\mu_{12}\Big),
		\label{vD1gen}\\[4pt]
		\langle D_2\lvert\rho_p\rvert B\rangle
		&=\frac{\sqrt2}{18}\,
		\Big(\mu_{12}e^{-i\alpha_2}+\mu_{12}^{*}e^{i\alpha_2}+\mu_{13}^{*}e^{i\alpha_3}	
		\notag\\[-2pt]
		&\hspace{2.0em}
		-2\mu_{13}e^{-i\alpha_3}+\mu_{23}^{*}e^{i(\alpha_3-\alpha_2)}\nonumber\\
		&\hspace{2.0em}
		-2\mu_{23}e^{i(\alpha_2-\alpha_3)}\Big).
		\label{vD2gen}
	\end{align}
	These give $C_{BD}^{\text{3-slit}}$ as a function of detector position through the $\alpha_i$. At the detector point matched to the source, $\theta_2=\delta_2,\ \theta_3=\delta_3$ (i.e.\ $\alpha_i=0$),
	Eqs.~\eqref{vD1gen}--\eqref{vD2gen} reduces to
	\begin{align}
		\langle D_1\lvert\rho_p\rvert B\rangle
		&=\frac{\sqrt6}{18}\Big[(x_{13}-x_{23})
		-i\,(2y_{12}+y_{13}-y_{23})\Big],
		\label{vD1match}\\[4pt]
		\langle D_2\lvert\rho_p\rvert B\rangle
		&=\frac{\sqrt2}{18}\Big[(2x_{12}-x_{13}-x_{23})
		-3i\,(y_{13}+y_{23})\Big],
		\label{vD2match}
	\end{align}
	where $\mu_{ij}=x_{ij}+iy_{ij}$.
	From Eqs.~\eqref{vD1match}--\eqref{vD2match}, $\mathbf{v}=0$ requires
	\begin{align}
		x_{13}
		&=x_{23},\quad 2x_{12}=x_{13}+x_{23},\nonumber\\
		&y_{13}=-y_{23},\quad 2y_{12}=y_{23}-y_{13}.
		\label{vanish}
	\end{align}
	For real overlaps the first two conditions force $x_{12}=x_{13}=x_{23}$: the bright--dark coherence vanishes if and only if
		the three pairwise overlaps are equal. Thus, at phase matched point, the coherence between the detector-coupled
	and detector-invisible sectors is therefore generated only by an asymmetry of the path-distinguishing memory. For the representative asymmetric configuration $\mu_{12}=a$, $\mu_{13}=\mu_{23}=b$ (real) one obtains
	the closed form
	\begin{equation}
		C_{BD}^{\text{3-slit}}=\frac{2\sqrt2}{9}\,\lvert a-b\rvert .
		\label{CBDclosed}
	\end{equation}
	
	It is instructive to contrast this with the 2-slit case. For
	$\rho_p = \begin{pmatrix} \rho_{11} & \Gamma_{12}\\ \Gamma_{12}^{*} & \rho_{22}\end{pmatrix}$
	in the path basis, the single bright--dark matrix element is
	\begin{equation}
		\langle D|\rho_p|B\rangle
		= \tfrac{1}{2}\left(\rho_{11}-\rho_{22}\right)
		+ i\,\mathrm{Im}\!\left(e^{-i\theta}\,\Gamma_{12}\right),
	\end{equation}
	so that
	\begin{equation}
		C_{BD}^{\text{2-Slit}}
		= \sqrt{\left(\rho_{11}-\rho_{22}\right)^{2}
			+ 4\,\mathrm{Im}^{2}\!\left(e^{-i\theta}\,\Gamma_{12}\right)}\, ,
	\end{equation}
	which for balanced populations reduces to
	$C_{BD}^{\text{2-slit}} = 2\,|\mathrm{Im}(e^{-i\theta}\Gamma_{12})|$; this is precisely the
	cross term of Eq.~\eqref{rho-dark-basis}. The two-path bright--dark coherence is therefore
	generically nonzero, but it is completely determined by the populations and the
	single complex coherence $\Gamma_{12}$: it oscillates with detector position, its
	maximum over the screen equals $2|\Gamma_{12}|$, and it vanishes at the
	phase-matched point whenever the overlap is real. It carries no information beyond
	the ordinary fringe visibility and phase.
	
	For $M \geq 3$ paths the situation is qualitatively different. Even at the
	phase-matched point ($\alpha_i = 0$) and for purely real overlaps---conditions
	under which $C_{BD}^{\text{2-slit}}$ vanishes identically---Eqs.~\eqref{vD1match}--\eqref{vD2match} show that for $M=3$, 3-slit bright dark coherence
	$C_{BD}^{\text{3-slit}}$ is generated by asymmetry among the pairwise overlap magnitudes,
	with the closed form of Eq.~\eqref{CBDclosed}. This mechanism has no two-path analogue, because
	with a single overlap $\mu_{12}$ the notion of an asymmetry across pairs does not
	exist. Thus, unlike single visibility parameter, $C_{BD}$ is  sensitive to relative structure among the $M(M-1)/2$ which-path overlaps.
	
Unlike $C_D$, the quantity $C_{BD}$ is directly tied to a measurement on the photon: a unitary acting on the path space that mixes $|B\rangle$ with the dark direction $\mathbf{v}/|\mathbf{v}|$ converts bright--dark coherence into detectable bright population, and the  population transfer achievable this way is set by $C_{BD}$. We emphasize that $C_{BD}$ does not govern recovery by ancilla postselection (the quantum eraser of Sec.~V.E), which acts on the memory rather than the photon and can revive interference even when $C_{BD}=0$. Because $C_{BD}$ is anchored to the physically privileged bright direction, it is invariant under the basis freedom of the dark frame.

	\section{Quantum eraser}
	\label{sec:eraser}
	
	The dark-sector measures $S_D$ and $C_{BD}$ characterize the photonic
	state after the which-path memory has been traced out. A quantum eraser instead measures the memory in a suitably chosen basis and postselects the outcome, preparing conditional photonic subensembles, following the general quantum-erasure principle introduced by Scully and Drühl\cite{ScullyDruhl1982}. We show that
	this operation distinguishes pure from mixed dark sectors in a way the
	bare detector response cannot. Further, it separates two
	physically different origins of dark-sector mixedness.

	For the marked single photon with a controlled memory given by \eqref{marked-state}, the three memory states span a subspace of the memory Hilbert space whose dimension equals the rank of the Gram matrix
	\begin{equation}
		G = 	\begin{pmatrix}
			1 & \mu_{12} & \mu_{13} \\
			\mu_{12}^{*} & 1 & \mu_{23} \\
			\mu_{13}^{*} & \mu_{23}^{*} & 1
		\end{pmatrix},
		\label{gram}
	\end{equation}
	with
	\begin{align}
		\det G	
		&= 1 + 2\,\mathrm{Re}\!\left(\mu_{12}\mu_{23}\mu_{31}\right)
		- |\mu_{12}|^2 \nonumber\\
		&- |\mu_{13}|^2 - |\mu_{23}|^2 .
		\label{detG}
	\end{align}
 For generic memory overlaps $\mu_{ij}$, $\det G \neq 0$, so $\mathrm{rank}\,G = 3$ and the which-path memory is a genuine qutrit, requiring a three-outcome memory basis.
	
	Let
	\begin{equation}
		\mathcal{H}_{M}
		=
		\operatorname{span}
		\left\{
		\ket{m_1},\ket{m_2},\ket{m_3}
		\right\}
	\end{equation}
	denote the Hilbert space of the memory states, and let
	\begin{equation}
		S=\left\{\ket{s_r}\right\}_{r=1}^{d},
		\qquad
		d=\dim\mathcal{H}_{M}
		=\operatorname{rank}G,
	\end{equation}
	be an arbitrary orthonormal measurement basis on this support.
	For the rank-three case to be considered below, \(r=1,2,3\).

	A projective measurement of the memory in the basis \(S\) gives outcome
	\(r\) with probability
	\begin{align}
		P_r
		&=
		\bra{\Psi_M}
		\left(
		I_p\otimes\ket{s_r}\bra{s_r}
		\right)
		\ket{\Psi_M}
		\nonumber\\
		&=
		\frac{1}{3}
		\sum_{i=1}^{3}
		\left|
		\langle s_r|m_i \rangle
		\right|^2.
		\label{memory_outcome_probability}
	\end{align}
	Conditioned on outcome \(r\), the photon is prepared in the pure state
	\begin{equation}
		\ket{\phi_r}
		=
		\frac{1}{\sqrt{3P_r}}
		\sum_{i=1}^{3}
		e^{-i\delta_i}
		\langle s_r|m_i \rangle\ket{i}.
		\label{conditional_photon_state}
	\end{equation}
	Equation~\eqref{conditional_photon_state} shows that an
	outcome-resolved memory measurement can localize path coherence in the
	conditional photon subensemble. Whenever at least two of the amplitudes
	\(\langle s_r|m_i \rangle\) are nonzero, the corresponding conditional state
	contains path coherence.
	
	The normalized path coherence of the conditional photon state is
	\begin{align}
		\mathcal{C}_{\mathrm{path}}
		\left(\ket{\phi_r}\right)
		&=
		\frac{1}{2}
		\sum_{i\neq j}
		\left|
		\bra{i}\phi_r\rangle
		\langle\phi_r\ket{j}
		\right|
		\nonumber\\
		&=
		\frac{
			\displaystyle
			\sum_{i<j}
			\left|\langle s_r|m_i \rangle\right|
			\left|\langle s_r|m_j \rangle\right|
		}{
			\displaystyle
			\sum_{i=1}^{3}
			\left|\langle s_r|m_i \rangle\right|^2
		}.
		\label{conditional_path_coherence}
	\end{align}
	Note that an arbitrary memory-measurement basis does not necessarily erase the
	which-path information. For example, if the memory states are mutually
	orthogonal and the measurement is performed in the memory basis,
	\begin{equation}
		\ket{s_r}=\ket{m_r},
	\end{equation}
	then
	\begin{equation}
		\ket{\phi_r}
		=
		e^{-i\delta_r}\ket{r},
		\qquad
		\mathcal{C}_{\mathrm{path}}
		\left(\ket{\phi_r}\right)=0.
	\end{equation}
	This measurement reveals the path record rather than erasing it.
	
	More generally, different choices of $S$ produce different
	conditional decompositions of the same reduced photon state. If the
	measurement outcomes are ignored, completeness of the measurement
	basis gives
	\begin{align}
		\sum_r P_r\ket{\phi_r}\bra{\phi_r}
		&=
		\frac{1}{3}
		\sum_{i,j=1}^{3}
		e^{-i(\delta_i-\delta_j)}
		\left[
		\sum_r
		\langle s_r|m_i\rangle
		\langle m_j|s_r\rangle
		\right]\nonumber\\
		&\hspace{4.0em}
		\times (\ket{i}\bra{j})
		\nonumber\\
		&=
		\frac{1}{3}
		\sum_{i,j=1}^{3}
		e^{-i(\delta_i-\delta_j)}
		\langle m_j|m_i \rangle
		\ket{i}\bra{j}
		\nonumber\\
		&=
		\rho_p.
		\label{unconditioned_reconstruction}
	\end{align}
	Thus, every memory measurement does not restore unconditional photon
	coherence. Coherence recovery is conditional on the selected memory basis and hence  restricted to postselected subensemble. We therefore refer to \(S=\{\ket{s_r}\}\) at this stage as a general memory-measurement basis. An erasure basis corresponds to measurement outcome  when it contains no accessible
	which-path information, namely when
	\begin{equation}
		\left|\langle s_r|m_1 \rangle\right|^2
		=
		\left|\langle s_r|m_2 \rangle\right|^2
		=
		\left|\langle s_r|m_3 \rangle\right|^2.
		\label{complete_erasure_condition}
	\end{equation}
	If this condition holds for every outcome of nonzero probability, then we call 	$S$ i a complete erasure basis. Otherwise, the measurement may produce partial erasure, complete erasure for selected outcomes, or no erasure at all.
	
	\subsection{Redistribution of bright and dark weights by memory postselection}
	We now discuss the effect of memory post-selection on the bright-dark decomposition.
	Consider the detector point that is perfectly dark in the unmarked
	experiment, where
	\begin{equation}
		\langle B|S'_3 \rangle=0.
	\end{equation}
	Before conditioning on a memory-measurement outcome, the marked state
	gives a residual bright weight
	\(W_B^{\mathrm{mark}}=\bra{B}\rho_p\ket{B}\), set by the overlaps
	\(\mu_{ij}\); for partial marking this is small but nonzero. The
	conditional bright weight in outcome $r$
	\begin{equation}
		W_B^{(r)}
		=
		\left|\langle B|\phi_r\rangle\right|^2,
	\end{equation}
	depends strongly on the memory basis. The memory basis can be chosen so that one outcome acquires the entire ensemble bright contribution within the six-parameter family, while the remaining outcomes are driven further into darkness. The average of the three outcomes reproduce the marked bright weight:
	\begin{equation}
		\sum_{r=1}^{3}P_rW_B^{(r)}
		=
		W_B^{\mathrm{mark}},
	\end{equation}
	which follows directly from the completeness relation:
	\begin{equation}
		\sum_{r=1}^{3}\ket{s_r}\bra{s_r}=\mathds{1}_{3}
		\end{equation}
	
	Define the memory-space vector
	\begin{equation}
		\ket{w}
		=
		\frac{1}{\sqrt{3}}
		\sum_i
		e^{-i\delta_i}
		\sqrt{3}\langle B|i \rangle
		\ket{m_i}
	\end{equation}
	so that
	\[
	W_B^{\mathrm{mark}}
	=
	\frac{1}{3}\langle w|w\rangle,
	\qquad
	P_rW_B^{(r)}
	=
	\frac{1}{3}
	\left|\langle s_r|w\rangle\right|^2.
	\]
	Choosing the memory basis such that
	\[
	\ket{s_1}
	=
	\frac{\ket{w}}{\lVert\ket{w}\rVert}
	\]
	therefore concentrates the entire marked bright weight into a single
	outcome,
	\begin{equation}
		W_B^{(1)}
		=
		\frac{W_B^{\mathrm{mark}}}{P_1}
		\geq W_B^{\mathrm{mark}},
		\qquad
		W_B^{(2)}=W_B^{(3)}=0,
	\end{equation}
	driving the two remaining outcomes exactly dark. For example, choose the asymmetric real qutrit overlaps
	\(\mu_{12}=a=1/2\) and
	\(\mu_{13}=\mu_{23}=b=1/4=a/2\) at the dark point
	\(\boldsymbol{\delta}=(0,2\pi/3,4\pi/3)\), with \(\theta_i=0\).
	The Gram determinant is \(\det G=11/16>0\), and one finds
	\begin{equation}
		W_B^{\mathrm{mark}}
		=
		\frac{2}{9},
		\qquad
		P_1
		=
		\frac{23}{96},
		\qquad
		W_B^{(1)}
		=
		\frac{64}{69},
	\end{equation}
	while \(W_B^{(2)}=W_B^{(3)}=0\). Thus, postselection concentrates
	the full marked bright weight into one subensemble and enhances the
	conditional detection probability by the  factor
	\begin{equation}
		\frac{W_B^{(1)}}{W_B^{\mathrm{mark}}}
		=
		\frac{96}{23}
	\end{equation}
	However, the postselected ensembles recompose the reduced photon state obtained
	after tracing over the memory:
	\begin{equation}
		\sum_{r=1}^{3}
		P_r\ket{\phi_r}\bra{\phi_r}
		=
		\operatorname{Tr}_M
		\left(
		\ket{\Psi_M}\bra{\Psi_M}
		\right)
		=
		\rho_p.
	\end{equation}
	Thus, the memory measurement redistributes the bright contamination introduced by which-path marking among the conditional subensembles: it is concentrated in the $s_1$
	outcome, while the $s_2$ and $s_3$  outcomes recover pure dark states with $W_B^{(2)}=W_B^{(3)}=0$. The original darkness is therefore recovered conditionally by postselection, while the ensemble-averaged bright weight remains unchanged. The memory measurement therefore does not create additional detection
	probability; it just redistributes the marked bright weight among the three
	conditional subensembles. This detector-specific bright revival is not necessarily a recovery of
	path coherence, because bright weight at a single detector point is not
	itself a coherence measure. Conditional coherence recovery must be
	established separately from the conditional photon state.
	
	\subsection{Conditional recovery of path coherence}
	\label{sec:assisted_coherence}
	A natural quantity considered in this context is the probability-weighted conditional dark-sector purity:
	
	\begin{equation}
		\bar P_D(S)=\sum_r P_r \text{Tr}(\rho_D^{(r)})^2,
		\end{equation}
	where $\rho_D^{(r)}$	is the normalized dark-sector state conditioned on memory outcome $r$. Although $\bar P_D$  quantifies how pure the post-selected dark components are, it does not by itself quantify quantum erasure or recovery of interference. In particular, $\bar P_D=1$ may occur even when the conditional photon states carry no multi-path coherence, for example when the memory measurement resolves the path and prepares pure single-path states.

To quantify the recoverability of path coherence (i.e the wave character)  by a memory measurement, we use the normalized $\ell_1$-norm coherence in the path basis. For a three-path density operator
	$\rho$, it is defined as
	\begin{equation}
		\mathcal{C}_{\mathrm{path}}(\rho)
		=\frac{1}{2}\sum_{i\neq j}\left|\rho_{ij}\right|
		=\sum_{i<j}\left|\rho_{ij}\right|,
		\qquad 0\leq \mathcal{C}_{\mathrm{path}}(\rho)\leq 1 .
		\label{Cpath_def}
	\end{equation}
 Equation~\eqref{Cpath_def} is defined in the physically fixed path basis and is therefore independent of any $U(2)$ rotation of dark frame.
	
	Consider marked photon state with controlled memory states $\{\ket{m_i}\}$ and (uncontrolled) environment states $\{\ket{e_i}\}$
	\begin{equation}
		|\Psi_{ME}\rangle
		=\frac{1}{\sqrt 3}\sum_{i=1}^{3}
		e^{-i\delta_i}|i\rangle|m_i\rangle|e_i\rangle ,
				\label{global_ME_CD}
		\end{equation}
	
	Tracing over  both memory and environment, the reduced photon state becomes
	\begin{equation}
		(\rho_p)_{ij}
		=\frac{1}{3}e^{-i(\delta_i-\delta_j)}
		\mu_{ji}\eta_{ji},
	\end{equation}
		where
	\begin{equation}
		\mu_{ij}=\langle m_i|m_j\rangle,\quad
		\eta_{ij}=\langle e_i|e_j\rangle,
	\end{equation}
	and hence
	\begin{equation}
		\mathcal{C}_{\mathrm{path}}(\rho_p)
		=\frac{1}{3}\sum_{i<j}|\mu_{ij}\eta_{ij}|.
		\label{unconditioned_CD}
	\end{equation}
	
	Let $S=\{|s_r\rangle \} $, where $r=1,2,3$; be a complete projective measurement
	of the accessible memory.  Outcome $r$  prepares the conditional photon state 
	
	\begin{equation}
		(\rho_P^{(r)})_{ij}=\frac{1}{3P_r}e^{-i(\delta_i-\delta_j)}\langle s_r|m_i\rangle \langle s_r|m_j\rangle^{*}\eta_{ji},
				\label{cond_rhop_r}
		\end{equation}
		where,
		\begin{equation}
		  P_r=\frac{1}{3}\sum_i |\langle s_r|m_i\rangle|^2,
		\end{equation}
	where outcome $r$ occurs with probability $P_r$.

	We define the probability-weighted conditional coherence
	\begin{equation}
		\overline{\mathcal C}_\mathrm{path}(S)
		=\sum_{r=1}^{3}P_r\,	\mathcal{C}_{\mathrm{path}}\!\left(\rho_p^{(r)}\right).
		\label{average_CD}
	\end{equation}
	Using outcome-$r$ conditioned reduced photon state given by Eq.~(\ref{cond_rhop_r}),  conditional coherence $\overline{\mathcal C}_\mathrm{path}(S)$ becomes
	\begin{equation}
		\overline{\mathcal C}_\mathrm{path}(S)
		=\frac{1}{3}\sum_{i<j}|\eta_{ij}|
		\sum_{r=1}^{3}
		\left|
		\langle s_r|m_i\rangle
		\langle s_r|m_j\rangle^{*}
		\right|.
		\label{average_CD_exact}
	\end{equation}
	This expression makes the roles of the two records---memory and environmental---explicit.  The memory
	measurement can recover the  coherence which was lost due to the accessible
	overlaps $\mu_{ij}$, whereas it cannot remove the factors $|\eta_{ij}|$
	imposed by the uncontrollable environment degrees of freedom.
	
	Every complete memory measurement satisfies
	\begin{equation}
		\mathcal C_\mathrm{path}(\rho_p)
		\leq \overline{\mathcal C}_\mathrm{path}(S)
		\leq \frac{1}{3}\sum_{i<j}|\eta_{ij}|.
		\label{CD_bounds}
	\end{equation}
	which follows from the completeness and the triangle inequality:
	\begin{align}
		\sum_r\left|
		\langle s_r|m_i\rangle
		\langle s_r|m_j\rangle^{*}
		\right|
		&\geq
		\left|\sum_r
		\langle s_r|m_i\rangle
		\langle s_r|m_j\rangle^{*}\right|
		=|\mu_{ij}|,
	\end{align}
	and the Cauchy--Schwarz inequality:
	\begin{align}
		\sum_r\left|
		\langle s_r|m_i\rangle
		\langle s_r|m_j\rangle^{*}
		\right|
		&\leq
		\sqrt{\sum_r|\langle s_r|m_i\rangle|^2}
		\sqrt{\sum_r|\langle s_r|m_j\rangle|^2}
		=1 .
	\end{align}

  For the normalised marker states $\{\ket{m_i}\}_{i=1}^{3}$ (spanning a space of dimension at most three), an orthonormal basis $\{\ket{s_r}\}_{r=1}^{3}$ can be chosen that carries no information about which of the three marker states was present. By suitable unitary rotation of the basis we get
	\begin{equation}
		|\langle s_r|m_1\rangle|^2
		=|\langle s_r|m_2\rangle|^2
		=|\langle s_r|m_3\rangle|^2
		\equiv p_r,
		\qquad \sum_r p_r=1.
		\label{path_unbiased_basis}
	\end{equation}
	Eq.(\ref{path_unbiased_basis}) can be obtained  by simultaneously choosing a zero diagonal for the two traceless Hermitian operators\cite{Horn1954,AuYeungPoon1979}
	$|m_1\rangle\langle m_1|-|m_2\rangle\langle m_2|$ and
	$|m_1\rangle\langle m_1|-|m_3\rangle\langle m_3|$.
	The upper equality in Eq.~\eqref{CD_bounds} is then attained pairwise,
	because each term becomes
	$|\langle s_r|m_i\rangle\langle s_r|m_j\rangle^*|=p_r$ and $\sum_r p_r=1$.
	Therefore the optimized memory-assisted coherence is
	\begin{equation}
			\mathcal C_\mathrm{path}^{\mathrm{opt}}
			=\frac{|\eta_{12}|+|\eta_{13}|+|\eta_{23}|}{3}
		\, ,
		\label{Cpath_opt}
	\end{equation}
	which corresponds to the memory basis which maximize $\overline{\mathcal C}_\mathrm{path}$.
	
	The coherence recovered by access to the memory is then
	\begin{equation}
			\Delta\mathcal C_\mathrm{path}
			=\mathcal C_\mathrm{path}^{\mathrm{opt}}-\mathcal C_\mathrm{path}(\rho_p)
			=\frac{1}{3}\sum_{i<j}
			|\eta_{ij}|\bigl(1-|\mu_{ij}|\bigr)
		\geq0 .
		\label{recoverable_CD}
	\end{equation}
	For the state given by Eq.~\eqref{global_ME_CD}, we may decompose coherence deficit $	(1-\mathcal C_\mathrm{path}(\rho_p))$  as
	\begin{equation}
		1-\mathcal C_\mathrm{path}(\rho_p)=\Delta\mathcal C_\mathrm{path}+
		(1-\mathcal C_\mathrm{path}^{\mathrm{opt}}),
		\label{Cpath_decomposition}
	\end{equation}
	where the first term on the right hand side is recoverable through suitable choice of the memory basis and the second term corresponds to irreducible coherence deficit due to environment. The second term should not be called a general measure of environmental
	mixedness.  Because it is defined in terms of optimized path coherence $C_\mathrm{path}^{\mathrm{opt}}$, it quantifies the coherence deficit that remains under the allowed memory-measurement protocol.
	
	With the decomposition \eqref{Cpath_decomposition}, it is natural to classify  limiting cases having differnt degree of mixedness that have a direct operational interpretation.
	\begin{enumerate}
		\item \textit{Memory-only record:} if $|\eta_{ij}|=1$ for all pairs, then
		\begin{equation}
			\mathcal C_\mathrm{path}^{\mathrm{opt}}=1,\qquad
			\Delta\mathcal C_\mathrm{path}=1-\mathcal C_\mathrm{path}(\rho_p).
		\end{equation}
	The apparent mixedness of the reduced photon state due to loss of coherence  by tracing over the accessible memory can be localized	in conditional photon subensembles with a suitable choice of memory basis $\{\ket{s_r}\}$.
		
		\item \textit{Environment-only record:} if $|\mu_{ij}|=1$ for all pairs,
		then
		\begin{equation}
			\mathcal C_\mathrm{path}^{\mathrm{opt}}=\mathcal C_\mathrm{path}(\rho_p),\qquad
			\Delta\mathcal C_\mathrm{path}=0.
		\end{equation}
		In this case measuring the memory the basis $\{\ket{s_r}\}$ cannot increase path coherence and maximum coherence obtained is given by \eqref{Cpath_opt}.
		
		\item \textit{Memory-Environment shared record:} when both $\mu$ and $\eta$ carry
		distinguishability, Eq.~\eqref{recoverable_CD} gives the part of the
		path-coherence deficit that can be recovered by measuring the memory. 
		
		\item \textit{Perfect records in both systems:} if
		$\mu_{ij}=\eta_{ij}=\delta_{ij}$, then
		\begin{equation}
			\mathcal C_\mathrm{path}(\rho_p)
			=\overline{\mathcal C}_\mathrm{path}(S)
			=\mathcal C_\mathrm{path}^{\mathrm{opt}}=0
		\end{equation}
		for every memory measurement. 
	\end{enumerate}

	We emphasize that this result has a limited scope.  The quantity $\mathcal C_{\mathrm{path}}^{\mathrm{opt}}$ measures optimized conditional path coherence, or equivalently the total pairwise interference capability of the postselected subensembles.  It does not decide the visibility at one fixed detector position. Equations~\eqref{Cpath_opt}--\eqref{Cpath_decomposition} hinges on the assumption of 
	balanced path amplitudes, product conditional records $|m_i\rangle|e_i\rangle$, and complete measurements of the accessible memory; their validity cannot be  extended  for arbitrary correlated memory--environment systems carrying path records.

	\section{Conclusion}
	\label{sec:conclusion}
	We studied the effect of a which-path memory ancilla on the three-slit interference pattern within the framework of the bright--dark state description.  Hilbert space structure is richer in case of multi-slit interference. In particular, three slits lead to one detector-coupled bright mode and a two-dimensional dark-mode subspace in the three-dimensional path space. Three slit interference pattern can be naturally explained in terms of bright and dark states. Principal maxima, secondary maxima, and minima correspond to complete, partial, and vanishing alignment, respectively, of the three path phasors with the detector-selected bright direction  
	
	We further studied the dark-subspace entropy and coherence. Whereas $C_D$ is not an intrinsic basis-independent characterization of the dark sector, bright-dark coherence $C_{BD}$ is basis independent as it is anchored to the detector-coupled bright mode. We found that, whereas two-slit bright-dark coherence $C_{BD}^{\text{2-slit}}$ carries no information beyond the ordinary fringe visibility, multi-path $C_{BD}$ for $M\geq3$ is sensitive also to the structure among $M(M-1)/2$ which-path overlaps. For $M=3$ case at the phase-matched detector point and for real overlaps, it vanishes only if pairwise overlaps are all equal.
	
	We finally discussed quantum erasure within ambit of bright-dark state formulation. We first discussed the effect of memory post-selection on the bright-dark decomposition. We found that, for a given detector position with a marked bright/dark weight,  suitable choice of memory outcomes distributes weights among themselves such that the  average reproduce the original  marked weight. This detector-specific bright revival is not necessarily a recovery of path coherence, because bright weight at a single detector point is not itself a coherence measure. We then asked how much path coherence can actually be recovered. We addressed that question with normalized $\ell_1$-norm coherence in the path basis. We found that every complete measurement basis satisfies inequality \eqref{CD_bounds}. For an normalised marker state $\{\ket{m_i}\}$ and a suitable choice of post-selected memory basis $\{\ket{s_r}\}$ satisfying \eqref{path_unbiased_basis} which carries no information about the marker states, one may decompose coherence deficit ($	(1-\mathcal C_\mathrm{path}(\rho_p))$  ) into recoverable (via choice of memory basis) and irreducible coherence deficit (due to uncontrollable) environment. However, this optimised coherence results have a limited scope; its validity hinges on the assumption of balanced path amplitudes, product conditional records and complete measurement of the accessible memory.
	
A natural direction for future work is to extend the BDS framework to higher-order interference phenomena following Mandel's framework for two-photon interference from independent atomic sources\cite{Mandel1964,Mandel1983,Ghosh1986}. Work in this direction is in progress and will appear somewhere else.

\bibliography{BDS_memory}
	
\end{document}